\documentclass[12pt]{JHEP3}
\usepackage{bm}
\usepackage{epsfig}
\usepackage{citesort}
\usepackage{graphicx}
\usepackage{multirow}

\usepackage{units}
\usepackage{epstopdf}
\usepackage{color}

\makeatletter

\providecommand{\tabularnewline}{\\}

\usepackage{subfig}

\title{ISW effect as probe of features in the expansion history of the Universe}

\author{Santanu Das\\
  Inter-University Centre for Astronomy and Astrophysics, Post
Bag 4, Ganeshkhind, Pune 411~007, India \\
E-mail: \email{santanud@iucaa.ernet.in}}

\author{ Arman Shafieloo\\
Asia Pacific Center for Theoretical Physics, Pohang, Gyeongbuk
790-784, Korea \\
$\&$\\
Department of Physics, POSTECH, Pohang, Gyeongbuk 790-784,
Korea \\
$\&$\\
Institute for the Early Universe, Ewha Womans University,
Seoul, 120-750, Korea\\
E-mail: \email{arman@apctp.org}}

\author{ Tarun Souradeep\\
Inter-University Centre for Astronomy and Astrophysics, Post
Bag 4, Ganeshkhind, Pune 411~007, India \\
E-mail: \email{tarun@iucaa.ernet.in}}

\abstract{
In this paper, using and implementing a new line of sight CMB code, called CMBAns~\cite{CMBAns}, that allows us to modify $H(z)$ for any given feature at any redshift we study the effect of changes in the expansion history of the Universe
on the CMB power spectrum. Motivated by the detailed analytical calculations of the effects of the changes in $H(z)$ on ISW plateau and CMB low multipoles, we study two phenomenological parametric form of the expansion history using WMAP data and through MCMC analysis. Our MCMC analysis shows that the standard $\Lambda$CDM cosmological model is consistent with the CMB data allowing the expansion history of the Universe vary around this model at different redshifts. However, our analysis also shows that a decaying dark energy model proposed in \cite{Shafieloo2009} has in fact a marginally better fit than the standard cosmological
constant model to CMB data. Concordance of our studies here with the previous analysis showing that Baryon Acoustic Oscillation (BAO) and supernovae data (SN Ia) also prefer mildly this decaying dark energy model to $\Lambda$CDM, makes this finding interesting and worth further investigation. 
}

\begin{document}
There are always theoretical degeneracies that makes it hard to distinguish
between different cosmological models. Though the standard spatially
flat $\Lambda$CDM model with power-law form of the primordial spectrum
provides a reasonably good fit to all cosmological observations using
a handful set of parameters, there is still space for some other models
to have a good concordance to the data. Using different cosmological
observations to probe a cosmological quantity is one of the best ways
we can approach this problem to break the degeneracies between different
models. One of the key problems in cosmology is to recover the dynamics
of the Universe and reconstruct the expansion history of the Universe
and so far supernovae type Ia (SN Ia) as standardised candles and
Baryon Acoustic Oscillation (BAO) data as standard rulers have been
the two main direct probes of the expansion history. It is also known
that changes in the expansion history of the Universe can affect the
angular power spectra of the Cosmic Microwave Background particularly
at low multipoles through Integrated Sachs Wolf (ISW) effect~\cite{ISW,KS85},
however little has been done in this direction~\cite{Linder2010,Linder2011,Linder2012,Linder2013}
due to complications of the analysis and indirect effect of the expansion
history on the CMB observables. In this paper we study the effect
of changes in the expansion history of the Universe on the CMB angular
power spectrum using a new CMB line of sight code called CMBAns~\cite{CMBAns}
developed and implemented to work on different forms of $H(z)$. The
advantage of this approach over using publicly available softwares such
as CAMB~\cite{CAMB} is that we can work on any desired form of $H(z)$
directly rather than considering different dark energy models with
equation of state of dark energy as an input. Considering perturbed
or un-perturbed dark energy, working on $H(z)$ would be in fact a
generalised study of different dark energy models which may have similar
effect on the expansion history of the Universe. After a detailed
analytical calculations of the effects of changes in $H(z)$ on CMB
low multipoles, we study two phenomenological models of the expansion
history of the Universe where both models include $\omega_{DE} = -1$ model
as a possibility through MCMC analysis and using WMAP CMB data~\cite{WMAP}.

We show that we can in fact use ISW effect to put reasonable constraints
on the expansion history of the Universe at distances that are beyond
the reach of supernovae or large scale structure data. We also show
that while standard $\Lambda$CDM model has a good concordance to
the data allowing $H(z)$ to vary around this model, a decaying dark energy model 
proposed in \cite{Shafieloo2009,decay2009,proceeding2009}
has also a very good fit to the data and in fact marginally better
than best fit $\Lambda$CDM model which makes this model interesting.

In the following first we go through some analytical calculations
and see how changes in the expansion history of the Universe can affect
the CMB low multipoles. Then we study two simple phenomenological
models of the expansion history where both of these models include
$\Lambda$CDM as a possibility in order to study how far we can deviate
from the standard model and still have a good fit to the data. Then
we present results and conclude.

\section{Problem formulation}

The CMB power spectrum is one of the most precisely measured quantities
in the theoretical astrophysics. There is no simple analytical expression
for calculating the CMB power spectrum with sufficient accuracy to
match the observations. An expression for calculating
the CMB temperature power spectrum \cite{Dodelson2003d,Peebles1994} can be written
as

\begin{equation}
C_{l}=\int_{0}^{\infty}|\Delta_{l}(k)|^{2}P(k)k^{2}dk\;.\label{eq:Cl}
\end{equation}

\noindent Here, $\Delta_{l}(k)$ is the brightness fluctuation function and
$P(k)$ is the primordial power spectrum.
The brightness fluctuation function can be written in terms of the
temperature source terms ($S_{T}(k,\tau)$) and the spherical Bessel
function ($j_{l}(x)$) of order $l$ as

\begin{equation}
\Delta_{l}(k)=\int_{0}^{\tau_{0}}S_{T}(k,\tau)\, j_{l}(k(\tau_{0}-\tau))d\tau\;,\label{eq:Brightness}
\end{equation}

\noindent where $\tau$ is the conformal time and $\tau_{0}$ represents the
conformal time at the present epoch i.e. at redshift $z=0$ and $k$
is the wave number. The exact expression for the temperature source
term in conformal gauge is given by

\begin{equation}
S_{T}(k,\tau)=g\left(\delta_{g}+\Psi-\frac{\dot{\theta}_{b}}{k^{2}}-\frac{\Pi}{4}-\frac{3\ddot{\Pi}}{4k^{2}}\right)+e^{-\mu}\left(\dot{\phi}+\dot{\Psi}\right)-\dot{g}\left(\frac{\theta_{b}}{k^{2}}+\frac{3\dot{\Pi}}{4k^{2}}\right)-\frac{3\ddot{g}\Pi}{4k^{2}}\;.\label{eq:Source_term}
\end{equation}

\noindent Here $\mu$ is the optical thickness at time $\tau$, $g$ is visibility
function and is given by $g=\dot{\mu}\exp(-\mu)$, $\delta_{g}$ is
the photon density fluctuation i.e. $\delta_{g}=\delta\rho_{g}/\rho_{g}$
where $\rho_{g}$ is the density of photons, $\theta_{b}=kv_{b}$,
where $v_{b}$ represents the velocity perturbation of the baryons,
$\phi$ and $\Psi$ are the metric perturbation variable where the
line element is given by $ds^{2}=a^{2}(\tau)\left\{ -\left(1+2\Psi\right)d\tau^{2}+\left(1-2\phi\right)dx^{i}dx_{i}\right\} $,
and $a(\tau)$is the scale factor. Here $i$'s run from $1$ to $3$.
$\Pi$ is the anisotropic stress and in most of the cases $\Pi$ and
its derivatives i.e. $\dot{\Pi}$ and $\ddot{\Pi}$ can be neglected
because they are small in comparison to the other terms. In all the
expressions overdot ($\dot{x}$) denotes the derivative with respect
to the conformal time.

The first term in the bracket in Eq.\ref{eq:Source_term} can be
interpreted in terms of the fluctuations in the gravitational potential
at the last scattering surface and is referred as the Sachs-Wolfe
(SW) term. The second term provides an integral over the perturbation
variables along the line of sight to the present era. This can be
interpreted in terms of variations in the gravitational potential
along the line of sight and this is often referred to as the Integrated
Sachs-Wolfe (ISW) term. The third term is known as the Doppler term
and that arises from the Doppler effect caused by the velocity perturbation
of the photons at the surface of last scattering.

The visibility function $g$ and its derivative $\dot{g}$ only peak
at the surface of the last scattering provided there is no re-ionization and in all the other parts it is zero. Therefore, the SW and the Doppler
term is only important at the surface of the last scattering. As the
ISW part is not multiplied with any such visibility function therefore
it is important throughout the expansion history. The ISW part can
be broken in two parts, 1) the ISW effect before the surface of last
scattering or the early ISW effect and 2) after the surface of last
scattering or the late ISW effect. Therefore, the total source term
can actually be broken into two independent parts, provided there
is no re-ionization,

\begin{equation}
S_{T}(k,\tau)=S_{T}^{Pri}(k,\tau)+S_{T}^{ISW}(k,\tau)\;.\label{eq:source_term_break}
\end{equation}

\noindent Here the $S_{T}^{Pri}(k,\tau)$, i.e. the primordial part consists
of the SW, Doppler and the early ISW part. The $S_{T}^{ISW}(k,\tau)$
part consists of the late time ISW part. As the dark energy only dominates
at the late time in the Universe therefore dark energy only affect
the ISW source term. 

The quantity we are interested in any CMB experiments is the angular
power spectrum, $C_{l}$ and Eq.\ref{eq:source_term_break} shows that
there are three independent terms in $C_{l}$,

\begin{equation}
C_{l}=C_{l}^{Pri}+C_{l}^{ISW}+2C_{l}^{Int}\;.\label{eq:Cl_break}
\end{equation}
 
The first term, which is

\begin{eqnarray}
C_{l}^{Pri} & = & \int_{0}^{\infty}k^{2}dk\left\{ \int_{0}^{\tau_{0}}\left[g\left(\delta_{g}+\Psi-\frac{\dot{\theta}_{b}}{k^{2}}\right)-\dot{g}\left(\frac{\theta_{b}}{k^{2}}\right)\right]\, j_{l}((\tau_{0}-\tau)k)d\tau\right.\nonumber \\
 &  & \left.+\int_{0}^{\tau_{*}}\left[e^{-\mu}\left(\dot{\phi}+\dot{\Psi}\right)\right]\, j_{l}((\tau_{0}-\tau)k)d\tau\right\} ^{2}\;,\label{eq:Cl_Pri}
\end{eqnarray}

\noindent is the contribution from pure SW, doppler effect and the early ISW.
This quantity is always positive since the integrand being a squared
term is positive. The second term

\begin{equation}
C_{l}^{ISW}=\int_{0}^{\infty}k^{2}dk\left\{ \int_{\tau_{*}}^{\tau_{0}}\left[e^{-\mu}\left(\dot{\phi}+\dot{\Psi}\right)\right]\, j_{l}((\tau_{0}-\tau)k)d\tau\right\} ^{2}\label{eq:Cl_ISW}
\end{equation}
 is the contribution from the late time ISW part. This part is also
positive because of the similar reason. As ISW effect is only important
at low multipoles, $C_{l}^{ISW}$ will provide a positive power at
low multipoles. $\phi$ and $\Psi$ are the perturbed gravitational
potential and they directly depend on the expansion history of the
Universe, i.e. $H(z)$. 

\begin{figure}
\includegraphics[width=0.9\columnwidth]{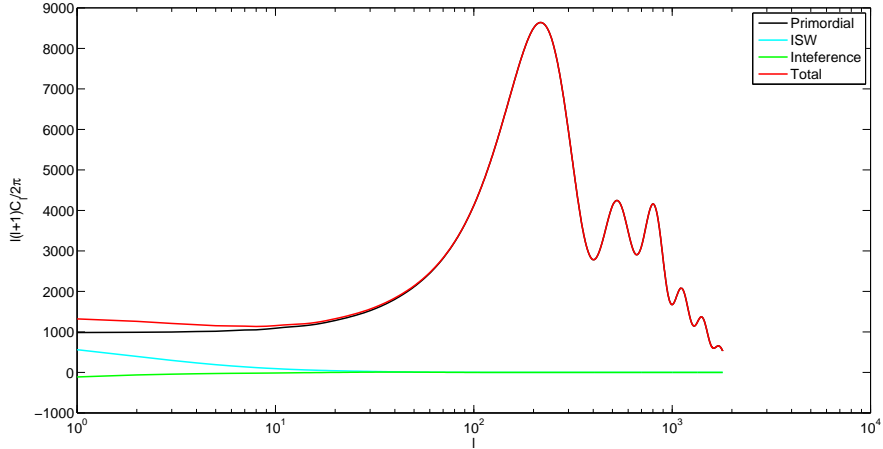}
\caption{\small Form of the angular power spectrum for $\Lambda$CDM model without the re-ionization and lensing effect. Black line represent the primordial part $C_{l}^{Pri}$, cyan line is for the ISW part $C_{l}^{ISW}$ and the green line shows the interference part $C_{l}^{Int}$. Red plot is showing the total angular power spectrum $C_{l}=C_{l}^{Pri}+C_{l}^{ISW}+2C_{l}^{Int}$. At low multipoles the power from the primordial part is almost constant. The increase in the power at low multipoles arises from the ISW and the interference part.}
\label{fig:Cl_LCDM}
\end{figure}


The third term 
\begin{equation}
C_{l}^{Int}=\int\Delta_{l}^{Pri}(k)\Delta_{l}^{ISW}(k)P(k)k^{2}dk\label{eq:Cl_Inte}
\end{equation}
 is the interference term between the primordial and ISW source terms.
The term $C_{l}^{Int}$ is important because unlike the other two
terms, $C_{l}^{Int}$ can either be positive or negative. If
we separate out the three terms then it can be seen that for $\Lambda$CDM
the interference term is actually negative. The interference
term in case of the $\Lambda$CDM model is very small as the two spherical
Bessel functions from the two independent parts are in general out of phase
and cancel each other. Therefore, the ISW term as a whole ($C_{l}^{ISW}+C_{l}^{Int}$)
typically increases the power at the low $C_{l}$ multipoles.


In Fig.\ref{fig:Cl_LCDM}, all three components of the angular power spectrum
for the $\Lambda$CDM model are shown independently. It can
be seen that the primordial part is almost flat at the low multipoles
and the increase of power at the low multipole is coming from the
ISW part. It is known that in case of the SCDM model the derivative
of the potentials are zero after the surface of the last scattering so no late time ISW effect exist there. It is also known that
the $H(z)$ varies faster in case of SCDM model then that of standard
$\Lambda$CDM model. So if we increase the $H(z)$ at the low redshift
then it is possible to decrease the power at the low CMB multipole.
Increasing the $H(z)$ slightly at the low redshift actually induces
two effects. First, it may decrease the power of the $C_{l}^{ISW}$
term and second it may make the $C_{l}^{Int}$ part more negative
and hence it may decrease the power at the low $C_{l}$ multipoles. 

Here it can also be noted that the ISW effect does not change the CMB
polarization power spectrum. The source term for the $E$ mode polarization
is given by 

\begin{equation}
S_{E}(k,\tau)=\frac{3}{16}\frac{g(\tau)\Pi(k,\tau)}{x^{2}}\;,
\end{equation}
where $x=k(\tau_{0}-\tau)$. As there is no potential dependent term, $E$ mode polarization source
term  remain unaffected by the ISW effect provided the distance
of the last scattering surface from the present era remains
fixed. So the polarization power spectrum i.e. $C_{l}^{ EE}$ will
remain fixed whereas the cross power spectrum i.e. $C_{l}^{TE}$ will
show some changes at low multipoles. 

In this work we have perturbed the Hubble parameter from the standard
$\Lambda$CDM model and calculated the angular power spectrum. It
may be noted that the $H(z)$ has been chosen to match the $\Lambda$CDM
model at the present redshift and the early epoch and deviation from
the $\Lambda$CDM model only occurs at some intermediate range. The details
of the models are discussed in the next section. Later on in this
paper we will also study a particular decaying dark energy model,
suggested in \cite{Shafieloo2009,decay2009,proceeding2009} where
at low redshifts $H(z)$ can have a larger values than $H(z)$ of
$\Lambda$CDM model that has been suggested by SN Ia and BAO data and
show that this model provides consistent fit to the CMB data. 

At this point, a brief description of CMBAns can be given as the following. CMBAns or Cosmic Microwave Background Anisotropy Numerical Simulation is a new CMB line of sight code written in the same line as that of CMBFAST~\cite{CMBFAST} and CAMB~\cite{CAMB} but with some extra features such as capability of computing CMB power spectra for different initial conditions, different inflationary models, extra dimensional models and also considering topology of the universe (ongoing project). CMBAns numerically integrates the linear metric and density perturbation equations as give in~\cite{ma_bert} over the conformal time to get the source terms and then convolve them with the Bessel functions to get the brightness fluctuation functions and then calculates the power spectrum by convolving the square of the brightness function with the primordial power spectrum. While writing the code we have taken special attention in the truncation conditions of the higher order photon and neutrino multipole moments to reduce the propagational errors. The code shows a very good agreement with CMBFAST and CAMB. With normal accuracy parameters the results fits with CAMB up to $0.2\%$ accuracy, and the accuracy increases with the enhanced accuracy boost parameters of CAMB and smaller grid size of CMBAns. One drawback of CMBAns over current version of CAMB is that CMBAns lensing module is based on~\cite{seljak96} and~\cite{zal_seljak98}, where as CAMB by default uses a lensing method based on~\cite{challinor_lewis05} which seems to be more accurate. However, for WMAP-7 likelihood this difference in the lensing module does not affect the results considerably. For this particular project the advantages of using CMBAns over CAMB is that in CMBAns we have a direct control over the expansion history, where in CAMB we need to write down the expansion history in terms of DE effective equation of state. In our analysis in some cases the DE effective energy density becomes negative and in such situations CAMB cannot calculate the expansion history. Also in CAMB, we cannot calculate the angular power spectrum when there is no dark energy perturbation. Therefore, in these cases instead of modifying CAMB it is easier and more straightforward to use CMBAns in which we have a full control over all the parameters.

It should be also noted that CMBAns does not calculate Primordial 
and ISW parts separately. In Fig.\ref{fig:Cl_LCDM} we have plotted these 
parts separately for no-reionization and no-lensing case just to give a 
physical understanding of the effects from different components. The results presented in the next section takes into account all effects including reionization and lensing.

\section{Changing the Expansion History}

As we have mentioned earlier, we study possible deviations in
$H(z)$ from the standard $\Lambda$CDM model with respect to the
CMB measured angular power spectrum from WMAP. We choose to put a
Gaussian deviation from the $H(z)$ expected from the standard $\Lambda$CDM
model at a particular redshift,

\begin{equation}
\frac{H(a)}{H_{\Lambda}(a)}=\left\{ 1+N\exp\left[-\left(\frac{a-a_{d}}{da}\right)^{2}\right]\right\} ^{\nicefrac{1}{2}}\label{eq:Hubble_parameter}
\end{equation}
 where $H_{\Lambda}(a)$ is the expansion history expected from $\Lambda$CDM
model, $a_{d}$ is the scale factor where we are putting the Gaussian
bump, $da$ is the width of the Gaussian bump and $N$ is the
amplitude of the Gaussian bump. 

There are two different ways to consider this perturbation in the
expansion history. Firstly, there can be different models such as
extra dimensional models of different scalar field models where the
dark energy can only change the expansion history of the Universe
and is not perturbed. There can be a second type of dark
energy models where the dark energy can itself gets perturbed. In 
these models not only dark energy affect the background expansion rate, but
its perturbation also directly changes the power spectrum.
In this paper we have analysed both types of these models.

\begin{figure}
\includegraphics[width=1\textwidth]{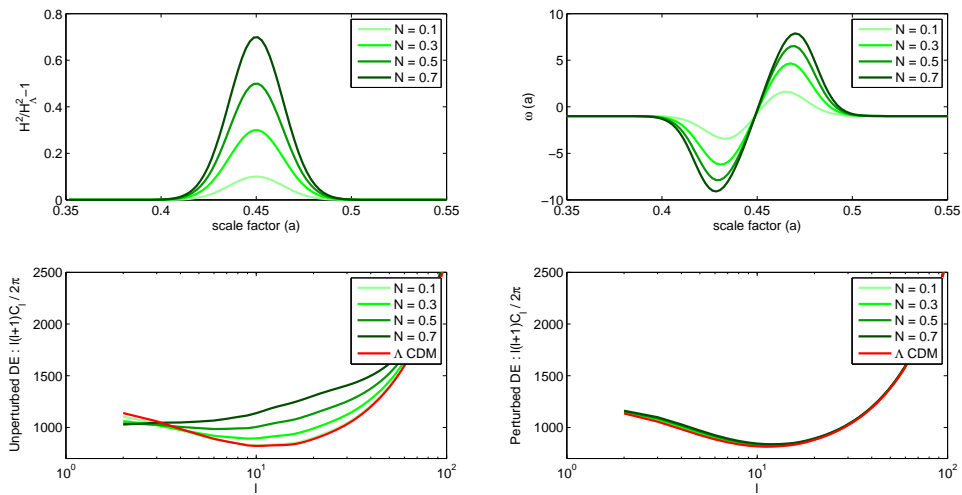}
\caption{\small Plots for different $N$ (amplitude of the Gaussian bump) using 
Eq.\protect\ref{eq:Hubble_parameter} for $H(z)$. The other two parameters are kept 
fixed at $a_{d}=0.45$ and $da=0.02$. Top-left : Plots of $\frac{H^{2}}{H_{\Lambda}^{2}}-1$ 
as a function of scale factor. Top-right : The equation of state of dark energy ($\omega(a)$) 
as a function of scale factor. Bottom-left : Angular power spectrum for unperturbed dark energy. 
Bottom-right: Angular power spectrum for perturbed dark energy.}
\label{fig:different_N}
\end{figure}

\begin{figure}
\includegraphics[width=1\textwidth]{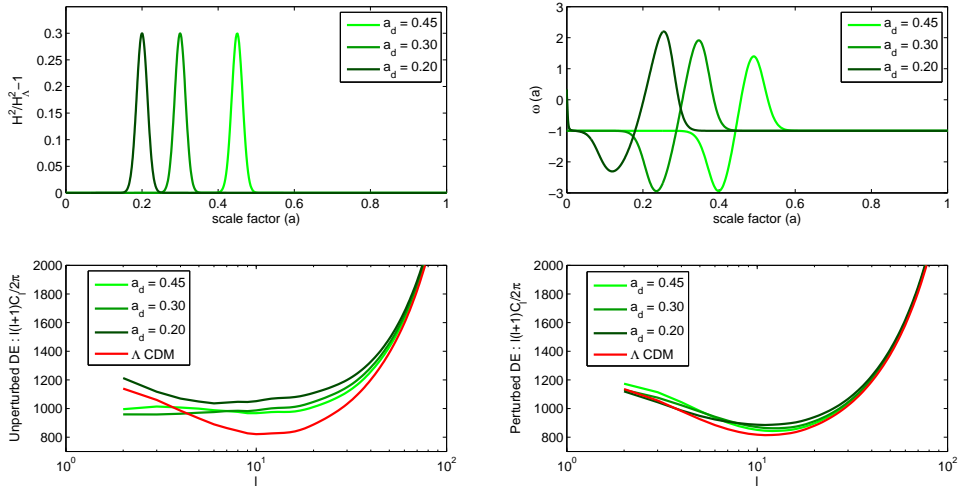}

\caption{\label{fig:different_ad} Plots for different $a_{d}$ (scale factor of the assumed Gaussian bump) using Eq.\protect\ref{eq:Hubble_parameter} for $H(z)$. The other two parameters are kept fixed at $N=0.3$ and $da=0.05$. Top-left : Plots of $\frac{H^{2}}{H_{\Lambda}^{2}}-1$
as a function of scale factor. Top-right : The equation of state of
dark energy ($\omega(a)$) as a function of scale factor. Bottom-left
: Angular power spectrum for unperturbed dark energy. Bottom-right
: Angular power spectrum for perturbed dark energy. }
\end{figure}

\begin{figure}
\includegraphics[width=1\textwidth]{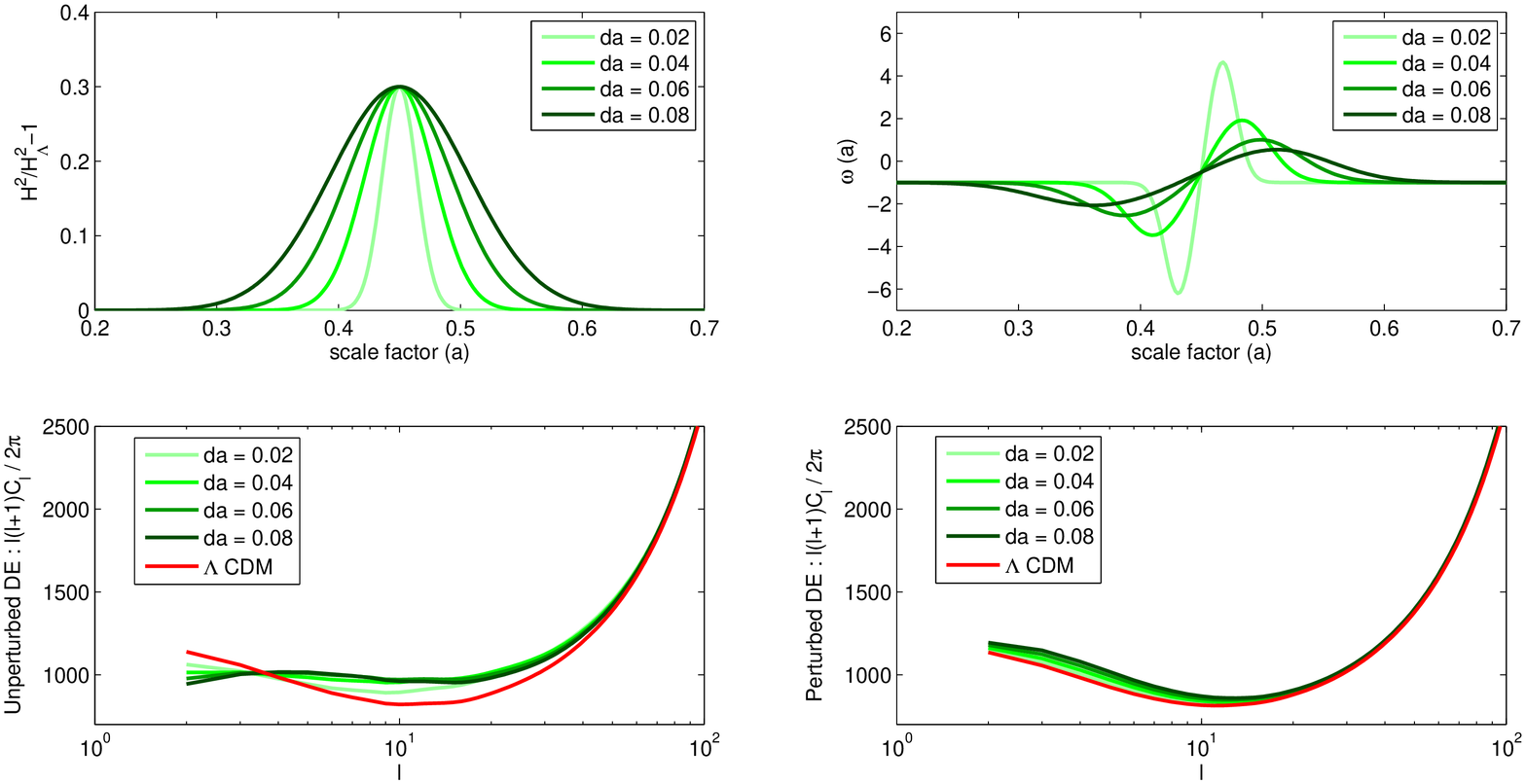}

\caption{\label{fig:different_da} Plots for different $da$ (width of the Gaussian bump) using Eq.\protect\ref{eq:Hubble_parameter}
for $H(z)$. The other two parameters are kept fixed at $a_{d}=0.45$
and $N=0.3$. Top-left : Plots of $\frac{H^{2}}{H_{\Lambda}^{2}}-1$
as a function of scale factor. Top-right : The equation of state of
dark energy ($\omega(a)$) as a function of scale factor. Bottom-left
: Angular power spectrum for unperturbed dark energy. Bottom-right
: Angular power spectrum for perturbed dark energy. }
\end{figure}

While $H(z)$ is given by Eq.\ref{eq:Hubble_parameter}, dark energy
equation of state and the $H(z)$ are related by the equation

\begin{equation}
w(a)=1-\frac{\frac{2}{3}\frac{a}{H(a)}\frac{dH(a)}{da}-\Omega_{0m}\frac{H_{0}^{2}}{H^{2}(a)}}{1-\Omega_{0m}\frac{a{}^{3}}{H^{2}(a)}H_{0}^{2}}\;.\label{eq:EoS}
\end{equation}
 Here we have considered that $\Omega_{r}$ is very small at the era
we are interested in and therefore the contribution from the radiation
part i.e. $\Omega_{r}$ can be neglected. $H_{0}$ is the Hubble parameter
at the present time and $\Omega_{0m}$ is the matter density, i.e. $\Omega_{0b}+\Omega_{0CDM}$
(also at the present era). In our case the expansion history can be written
as

\begin{equation}
\frac{H^{2}(a)}{H_{0}^{2}}=\left(\frac{\Omega_{0m}}{a^{3}}+\Omega_{\Lambda}\right)\left(1+N\exp\left(-\left(\frac{a-a_{d}}{da}\right)^{2}\right)\right)\;,\label{eq:HaoH0}
\end{equation}

\noindent and after few simple algebraic manipulation one can show that

\begin{equation}
\frac{1}{H^{2}(a)}\frac{dH^{2}(a)}{da}=-\left[\frac{\frac{3\Omega_{0m}}{a^{2}}}{\frac{\Omega_{0m}}{a^{3}}+\Omega_{\Lambda}}+\left(\frac{2}{da}\right)\left(\frac{a-a_{d}}{da}\right)\frac{N\exp\left(-\left(\frac{a-a_{d}}{da}\right)^{2}\right)}{1+N\exp\left(-\left(\frac{a-a_{d}}{da}\right)^{2}\right)}\right]\;.\label{eq:dhaoHada}
\end{equation}

Using Eq.\ref{eq:EoS}, Eq.\ref{eq:HaoH0} and Eq.\ref{eq:dhaoHada}
we can find out $\omega(a)$ as a function of the scale factor. The
perturbation equation which we have used for the dark energy \cite{Hannestad2005c,Weller2003}are
given by

\begin{equation}
\dot{\delta}_{x}=-3H\left(c_{s}^{2}-\omega_{x}\right)\left(\delta_{x}+3H\left(1+\omega_{x}\right)\frac{\theta_{x}}{k^{2}}\right)-\frac{3H\dot{\omega_{x}}\theta_{x}}{k^{2}}-\left(1+\omega_{x}\right)\theta_{x}-3\left(1+\omega_{x}\right)\dot{h}\label{eq:density-perturbation}
\end{equation}
 and 
\begin{equation}
\frac{\dot{\theta}_{x}}{k^{2}}=-H\left(1-3c_{s}^{2}\right)\frac{\theta_{x}}{k^{2}}+\Psi+c_{s}^{2}\delta_{x}/\left(1+\omega_{x}\right)\;.\label{eq:velocity-perturbation}
\end{equation}

\noindent Here $c_{s}^{2}=\frac{\delta P_{x}}{\delta\rho_{x}}$ and $c_{s}^{2}$
is kept to be unity in the analysis. $\delta_{x}$ is the density
perturbation of the dark energy and $\theta_{x}=ku_{x}$, where $k$
is the wave number and $u_{x}$ is the velocity perturbation of the
dark energy. $\dot{h}=\dot{\left(\frac{\delta a}{a}\right)}$ where the
over-dot denotes the derivative with respect to the conformal
time.

In Fig.\ref{fig:different_N} we show the computed power spectrum
for different $N$ values but constant $a_{d}$ and $da$ cases. Fig.\ref{fig:different_ad} shows the plots for variable $a_{d}$ and
the Fig.\ref{fig:different_da} shows plots for variable
$da$. Except the bump parameters all other parameters for the model
are same as that of the standard $\Lambda$CDM model. Plots are made
for both perturbed and unperturbed dark energy models and the
plot for the standard $\Lambda$CDM is also shown for comparison. One can notice that the unperturbed dark energy has more
effect on the low CMB multipoles through ISW effect (in comparison
with perturbed dark energy).

Here one should note that we have allowed for both positive and 
negative values of $N$. It is possible that for some negative values of $N$ the effective DE density becomes negative. This in fact might happen when $h^2(z) < \Omega_{0m}(1+z)^3$ where $h^2(z)=\frac{H^2(z)}{H_{0}^2}$. In the unperturbed DE case we have directly modified the $H(z)$ without considering any particular restriction for the underlying dark energy model. One should realise that $H(z)$ is directly related to the dynamical geometry of the Universe and the expansion history is governed by matter and the unknown dark energy which is not necessarily a scalar field. In some special cases of brane cosmological models (where the presence of the higher dimensions affect the expansion history), while $H(z)$ has a proper behaviour, the effective equation of state can have singularity and we can have effective negative density~\cite{Sahni_Shtanov2003,Shafieloo2006}.
However, in case of dark energy perturbation, it is not allowed to have a negative density for DE. In such cases where combination of our parameters resulted to negative DE density at some particular redshifts, DE density is set to be zero. The expansion history and the dark energy perturbation are also calculated accordingly.

\section{Results and Discussion}

\subsection{Bump Model}

The parametric form of $H(z)$ we use in this analysis given by Eq.\ref{eq:Hubble_parameter}
allows us to study how far we can deviate from the standard $\Lambda$CDM
model at different redshifts while still keeping the concordance to
the CMB data. This parametric form can also give us a hint if a particular
smooth deviation from the expansion history given by the standard
model may rise to a better fit to the CMB data and hence can be used
to test the consistency of the standard $\Lambda$CDM model to the
CMB observations.



In this analysis, we have used a numerical package called CMBAns
\cite{CMBAns}. For unperturbed dark energy models
the $H(z)$ has been changed directly and its effects are analysed.
In the case of perturbed dark energy, along with the modification
of the $H(z)$, the dark energy velocity and density perturbations
are included with the other perturbation equations. These dark energy
perturbations change the potential at the late time and hence also
influence the ISW effect.

For finding out the best fit set of parameters, a MCMC code using
the global metropolis algorithm has been used. The MCMC code uses
CMBAns \cite{CMBAns} for calculating the theoretical power spectra (temperature and polarization)
and the WMAP likelihood \cite{WMAP_likelihood} code for calculating
the likelihood of the theoretical power spectrum using the WMAP 7
year data~\cite{WMAP}. We have chosen a flat prior for all cosmological
parameters.

\begin{figure}
\subfloat[\label{fig:Unperturbed_DE_9}Two dimensional likelihood contours for the set
of 9 cosmological parameters]{\includegraphics[width=0.9\columnwidth]{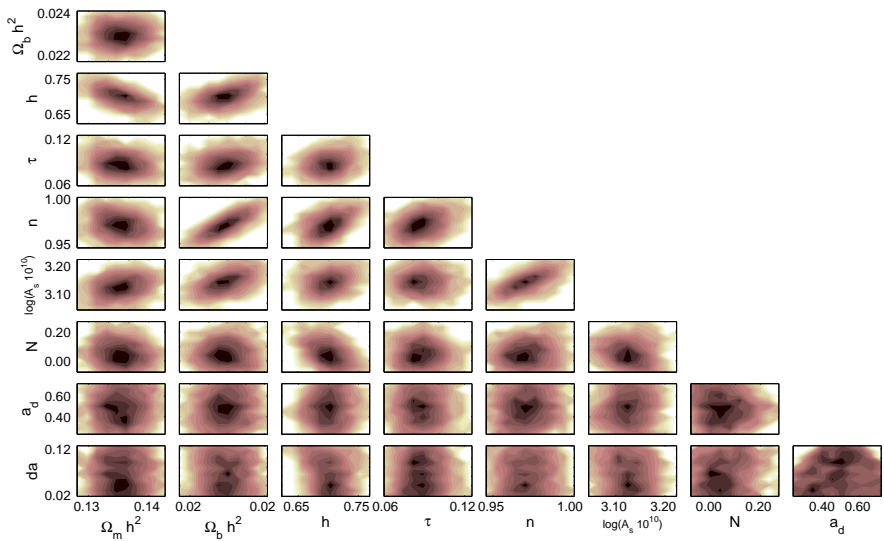}}

\subfloat[\label{fig:Unperturbed_DE_91D}One dimensional marginalized probability
distribution ]{\includegraphics[width=0.9\columnwidth]{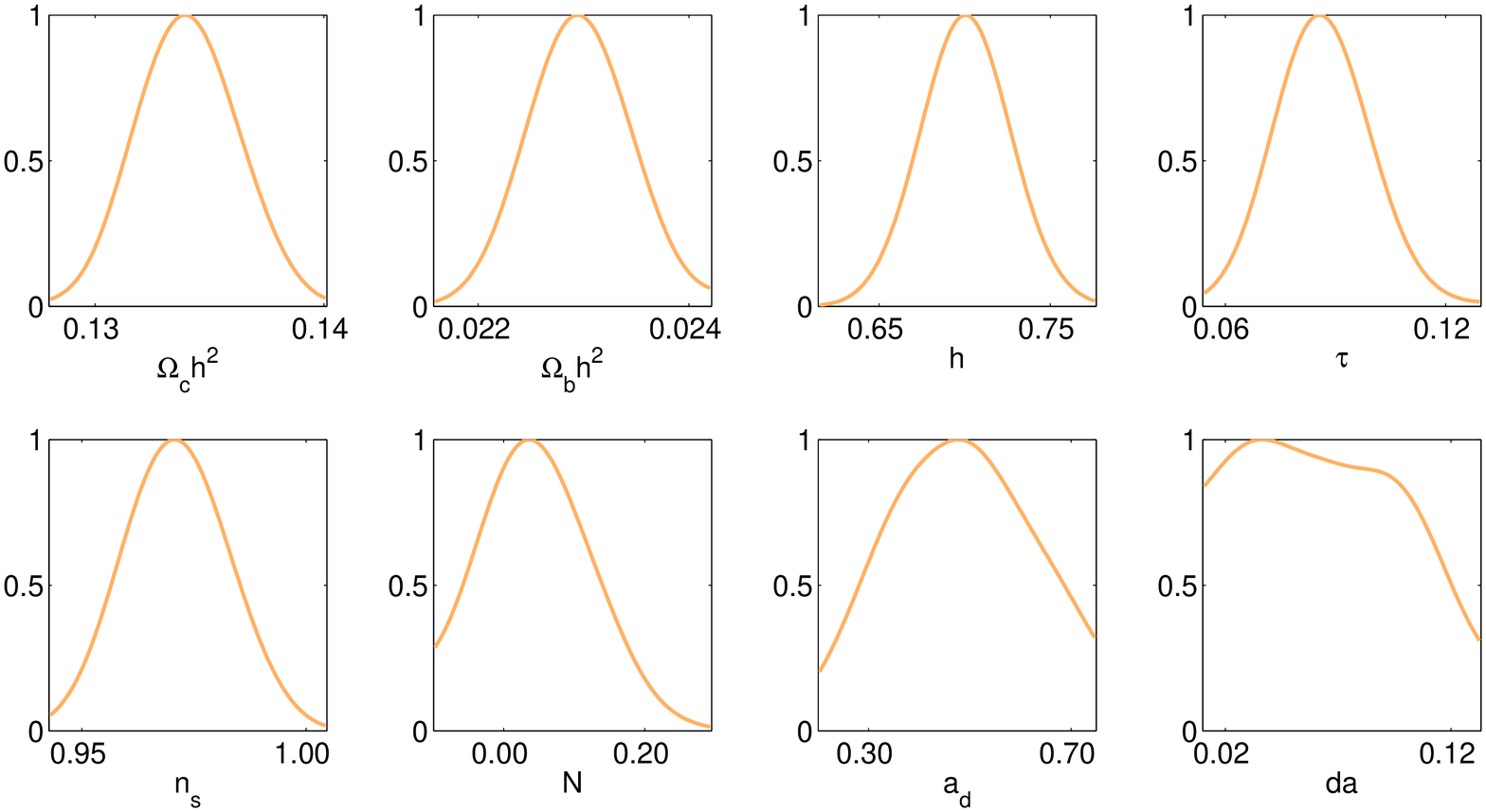}

}

\caption{Cosmological parameter estimation (9 parameters) assuming no perturbation for dark energy. $H(z)$ for the model is given by Eq.\protect\ref{eq:Hubble_parameter}. }
\end{figure}

Four set of analysis have been carried out with the Gaussian deviation
in the $H(z)$. The first set uses a 9 parameter set amongst which
6 are the standard cosmological parameters, namely baryon density
$\Omega_{b}h^{2}$, total matter density $\Omega_{0m}h^{2}$, $h$,
the Hubble parameter in a units of $100\, km/s/Mpc$, re-ionization
optical depth ($\tau$), scalar spectral index ($n_{s}$) and the
primordial power spectrum amplitude ($A_{s}$). Apart from
these 6 standard parameters we have used three extra parameters for
modifying the $H(z)$ using a Gaussian bump. These are the amplitude
of the Gaussian bump $N$, the red shift or scale factor $a_{d}$ at which
the Gaussian bump is placed and the width/standard deviation of the
Gaussian bump $da$. In Fig.\ref{fig:Unperturbed_DE_9} we
have shown the two dimensional likelihood of the parameters from the
MCMC analysis. The plots show that Gaussian parameters are almost
uncorrelated with all the cosmological parameters except the $H_{0}$
($h$) and $A_{s}$. There is a mild negative correlation between
$h$ and $N$ with correlation coefficient $~0.32$. As the Gaussian
bump parameters are very much uncorrelated with the standard model
parameters therefore we can expect all the parameters to be very much
close to the standard model parameters. The one dimensional marginalized
probability distribution for the 8 parameters are shown in Fig.\ref{fig:Unperturbed_DE_91D}. It shows that the distribution for
$da$ is almost flat for a wide range of values. Therefore the convergence
of the Gelman Rubin statistics 
for $da$ is very slow. 
The average values of the parameters and their standard deviation
are given in the table \ref{tab:Likelihood}. The results show
that the expansion rate of the Universe can deviate up to $10\%$
from the $H(z)$ given by $\Lambda$CDM model at some redshifts however,
the data seems to be clearly consistent with the $\Lambda$CDM model
and in fact adding these features to $H(z)$ seems not to significantly impact the fit to the data. Another noticeable result is that $h$ is
allowed to have a smaller values than those expected from the $\Lambda$CDM
model.

\begin{figure}
\centering
\begin{tabular}{lr}
\subfloat[\label{fig:Unparturbed32D}Two dimensional likelihood contours for the set of 4 
parameters]{\includegraphics[width=0.45\textwidth]{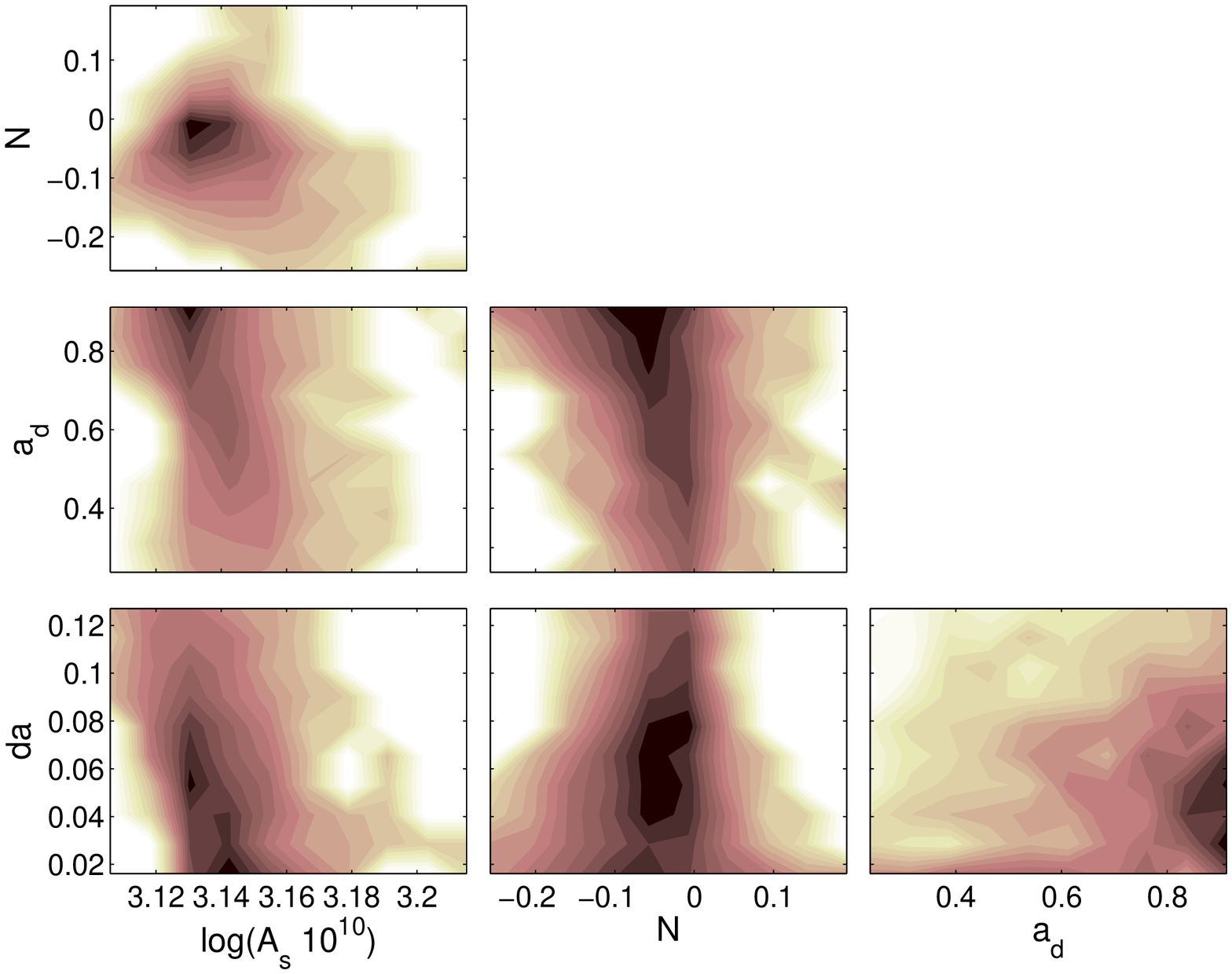}}
&
\subfloat[\label{fig:Unparturbed31D}One dimensional marginalized probability distribution]{\includegraphics[width=0.45\textwidth]{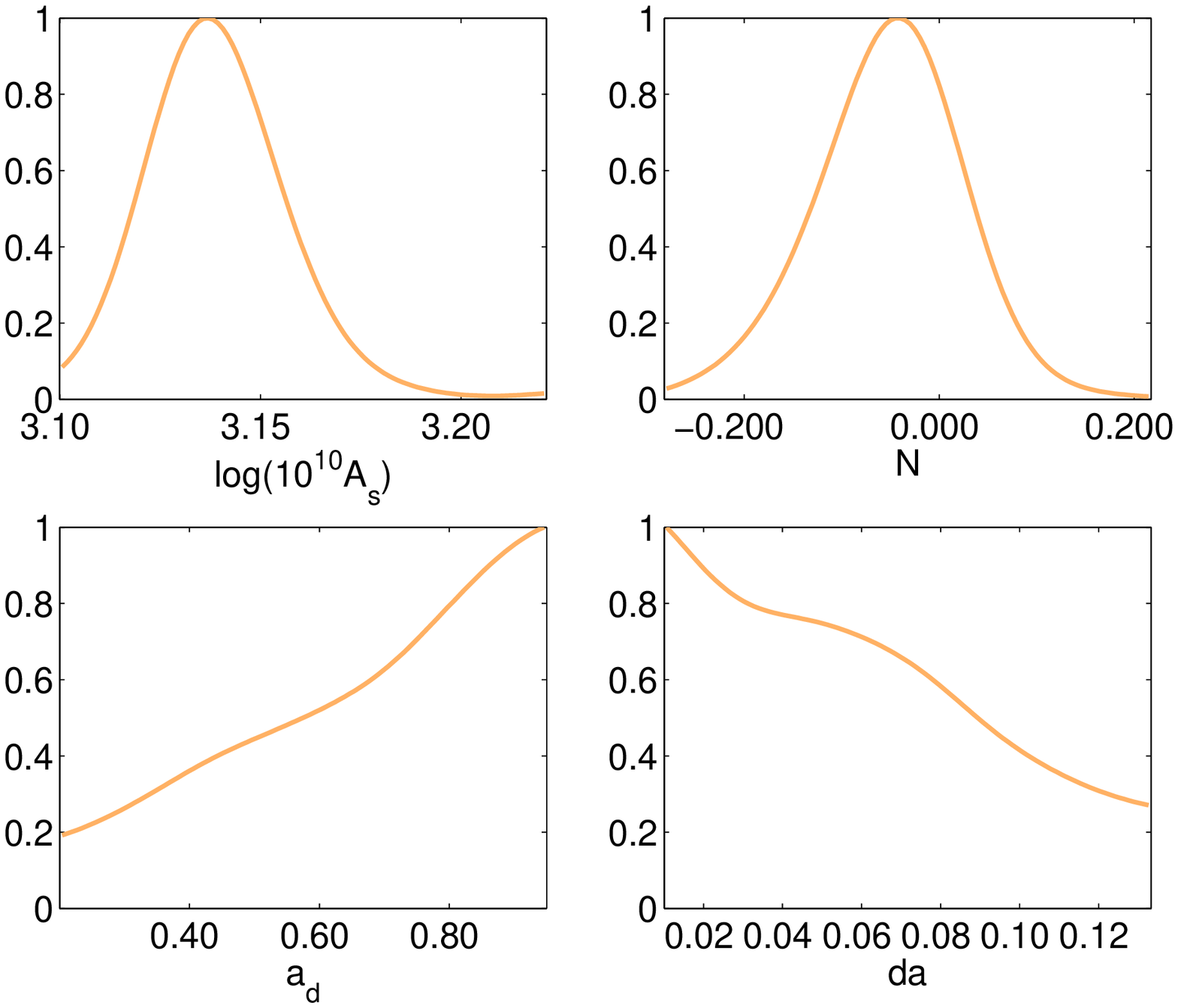}}\\
\end{tabular}
\caption{\label{fig:Unperturbed_DE_3} Parameter estimation for the bump model assuming no perturbation for dark energy. $H(z)$ for the model is given by Eq.\protect\ref{eq:Hubble_parameter} and other cosmological parameters are fixed at their best fit $\Lambda$CDM values.}
\end{figure}

The second analysis is carried out with a 4 parameter set. In this
case we keep all the standard cosmological parameters,
except $A_{s}$, fixed at their best fit $\Lambda$CDM model values. The parameters
which we use as the free parameters here are the parameters of the Gaussian bump, 
i.e. $N$, $a_{d}$ and $da$. As $A_{s}$ scales the power spectrum, it is important 
to make it a free parameter. Results are shown in Fig.\ref{fig:Unperturbed_DE_3}.
In Fig.\ref{fig:Unparturbed32D} we have plotted the two dimensional
likelihood contours for the set of 4 parameters. The plots show
that none of the two parameters are correlated. In Fig.\ref{fig:Unparturbed31D}
the one dimensional marginalized probability distributions are plotted.
The plots show that if all the standard cosmological
parameters are fixed to their standard model values, the constraints
on the bump parameters become tighter. However, height of the Gaussian
bump indicates that still we can have significant deviation from
$H(z)$ given by $\Lambda$CDM at some intermediate redshifts. The
width of the Gaussian bump also becomes narrower not allowing to deviate
from $\Lambda$CDM model continuously in a broad range. This is something
expected as we have fixed all the other parameters in the analysis
so the constraints on the remaining ones must become statistically
tighter. 
The values of the fitted parameters are tabulated in table \ref{tab:Likelihood}.


\begin{figure}
\subfloat[\label{fig:Perturbed_DE_9_2D}Two dimensional likelihood contours for the set
of 9 cosmological parameters]{\includegraphics[width=0.9\columnwidth]{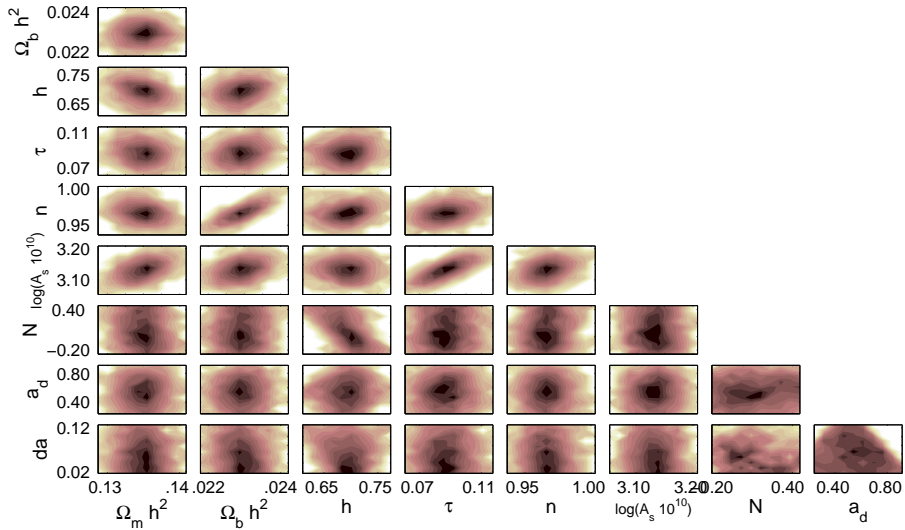}

}

\subfloat[\label{fig:Perturbed_DE_91D}One dimensional marginalized probability
distribution ]{\includegraphics[width=0.9\columnwidth]{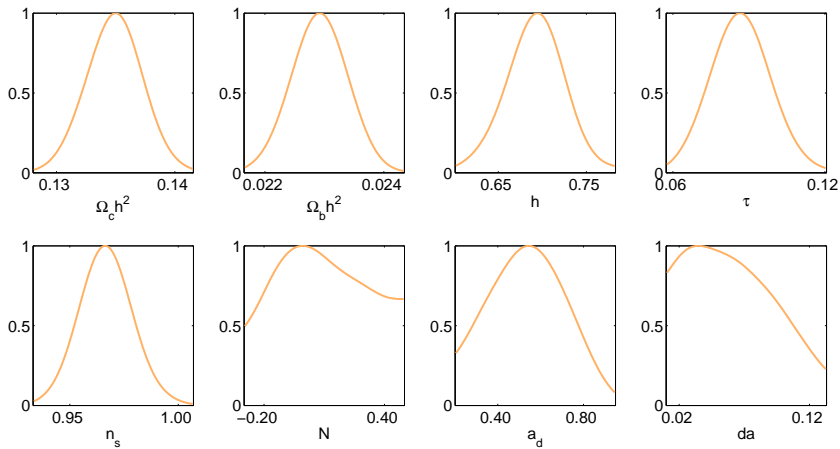}

}

\caption{\label{fig:Perturbed_DE_9}Cosmological parameter estimation (9 parameters) assuming perturbed dark energy. $H(z)$ for the model is given by Eq.\protect\ref{eq:Hubble_parameter}.}
\end{figure}

The third analysis is carried out with 9 parameters but with the perturbed
dark energy model. The parameter set is same as that of the first
parameter set. The results are shown in Fig.\ref{fig:Perturbed_DE_9}.
The two dimensional likelihood plots are shown in Fig.\ref{fig:Perturbed_DE_9_2D}
and the one dimensional marginalized probability distribution for the
8 parameters are shown in Fig.\ref{fig:Perturbed_DE_91D}. These
plots also show that there is a correlation between the Hubble parameter and 
the bump amplitude. The insertion of the bump in $H(z)$ changes the distance to the last scattering surface. The distance to last scattering surface is also sensitive to the Hubble parameter and this can explain the correlation between the bump amplitude and the Hubble parameter. 
Apart from this the bump parameters are
very much uncorrelated with all the other cosmological parameters.
Here, it can be seen that the distribution of $N$ and $da$ are
very much flat. The standard deviations of these parameters also reflect the 
same. The standard deviation for $N$ is larger in this
case in comparison to the previous unperturbed DE case. This allows
up to $20\%$ deviation from the expansion history given by $\Lambda$CDM
model at some intermediate redshifts around $z\approx0.6$. 
The average values of the fitted parameters are tabulated in table \ref{tab:Likelihood}
.

\begin{figure}
\centering
\begin{tabular}{lr}
\subfloat[\label{fig:Parturbed_42D}Two dimensional likelihood contours for the set of 4 parameters]{\includegraphics[width=0.41\columnwidth]{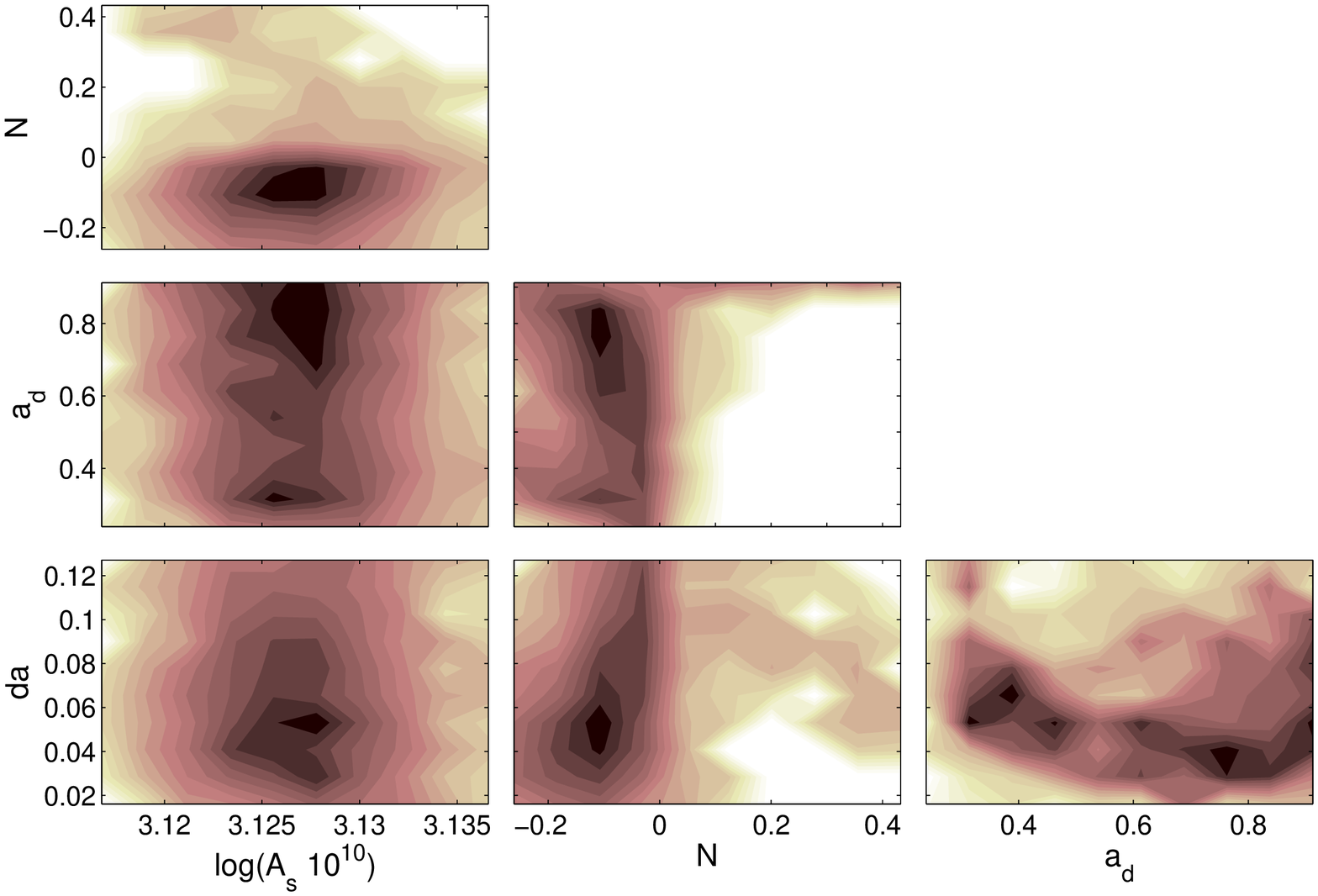}

}
&
\subfloat[\label{fig:Perturbed_41D}One dimensional marginalized probability distribution]{\includegraphics[width=0.41\columnwidth]{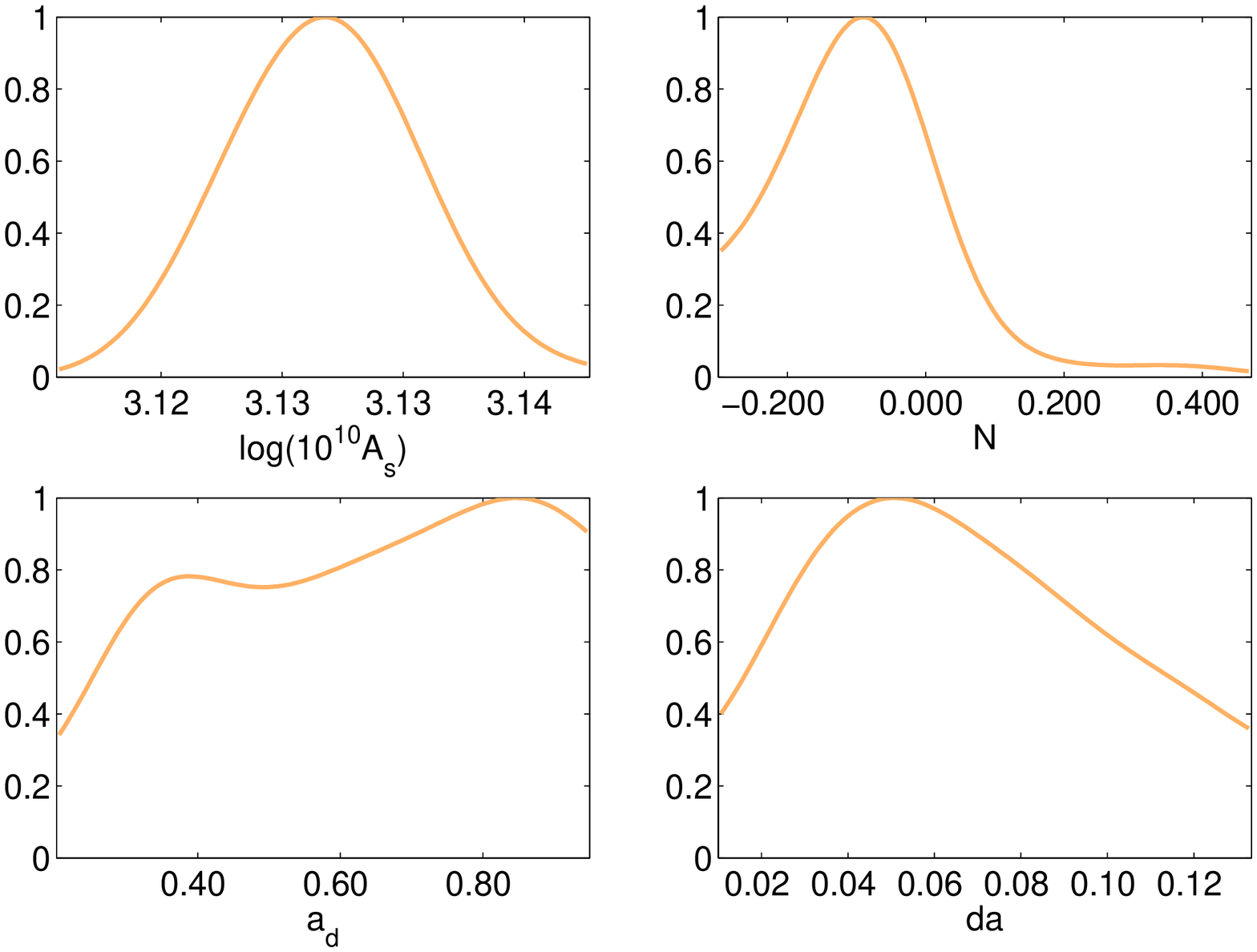}

}\\
\end{tabular}
\caption{\label{fig:Perturbed_DE_3}Parameter estimation for the bump model assuming perturbation for dark energy. $H(z)$ for the model is given by Eq.\protect\ref{eq:Hubble_parameter} and other cosmological parameters are fixed at their best fit $\Lambda$CDM values.}
\end{figure}

In the fourth case, we have analysed the perturbed dark energy model with
a 4 parameter set. All the standard cosmological parameters expect
$A_{s}$ are kept fixed. The two dimensional likelihood plots are shown
in Fig.\ref{fig:Parturbed_42D} and the one dimensional marginalized
probability distribution for the parameters are shown in Fig.\ref{fig:Perturbed_41D}.
The plots show that $a_{d}$ is following some double humped kind
of distribution. The average values of the fitted parameters are
tabulated in table \ref{tab:Likelihood}. These results show that
while we can in fact deviate the expansion history of the Universe significantly from the 
the standard $\Lambda$CDM model at some
intermediate redshifts and still having a good fit to the data, but
having the standard model close to centre of confidence contours indicates
toward robust consistency of the standard model to the data. Getting
only mild improvement in the likelihood to the data assuming few more
degrees of freedom, hints towards the fact that any Bayesian analysis
would still favour $\Lambda$CDM to the alternative ones.




\begin{center}
\begin{table}
\footnotesize
\begin{tabular}{|c|c|c|c|c|}

\hline 
 & Unperturbed (9 par)  & Unperturbed (4 par)  & Perturbed (9 par)  & Perturbed (4 par)\tabularnewline
\hline 
\hline 
$\Omega_{b}h^{2}$  & 0.0224$\pm$ 0.0004  & 0.0224  & 0.0224$\pm$0.0004  & 0.0224\tabularnewline
\hline 
$\Omega_{0m}h^{2}$  & 0.1332$\pm$0.0038  & 0.1336  & 0.1348$\pm$0.0040  & 0.1336\tabularnewline
\hline 
$h$  & 0.6997$\pm$0.0201  & 0.705  & 0.6897$\pm$0.0268  & 0.705\tabularnewline
\hline 
$\tau$  & 0.0865$\pm$0.0108  & 0.0848  & 0.0869$\pm$0.0103  & 0.0848\tabularnewline
\hline 
$n_{s}$  & 0.9708$\pm$0.0114  & 0.968  & 0.9670$\pm$0.0099  & 0.968\tabularnewline
\hline 
$\log(10^{10}A_{s})$  & 3.1254$\pm$0.0281  & 3.1421$\pm$0.0041  & 3.1319$\pm$0.0251  & 3.1252$\pm$0.0014\tabularnewline
\hline 
$N$  & 0.0478$\pm$0.0624  & -0.0523$\pm$0.0225  & 0.1215$\pm$0.2030  & -0.1085$\pm$0.0296\tabularnewline
\hline 
$a_{d}$  & 0.4798$\pm$0.1297  & 0.5938$\pm$0.0745  & 0.5938$\pm$0.0745  & 0.5389$\pm$0.0724\tabularnewline
\hline 
$da$  & 0.0617$\pm$0.0316  & 0.0505$\pm$0.018  & 0.0594$\pm$0.0298  & 0.0606$\pm$0.0116\tabularnewline
\hline 
$\Delta\chi^{2}$  & 0.3  & 0.2  & 1.4  & 0.7 \tabularnewline
\hline 
\end{tabular}\caption{\label{tab:Likelihood} Estimated cosmological parameters for the bump model. $H(z)$ for the model is given by Eq.\protect\ref{eq:Hubble_parameter}. Results are presented for both perturbed and unperturbed dark energy. Note that the $\Delta\chi^{2}$ values given here are for the best fit points in the parameter space.}
\end{table}

\end{center}

\subsection{Decaying Dark Energy Model}

Another set of analysis have been carried out using a parametric form
of the dark energy equation of state that allows the expansion of
the Universe undergo a slowing down in its acceleration at low redshifts.
The particular equation of state has been analysed by Shafieloo et.al.
in \cite{Shafieloo2009,decay2009,proceeding2009} to show that slowing
down of the acceleration in the expansion history of the Universe
which might come from a decaying dark energy model, can in fact improve
the fit to both supernovae and BAO data.

The dark energy equation of state is here taken as 
\begin{equation}
\omega(z)=-\left[1+\tanh((z-z_{t})\Delta z)\right]/2.
\label{eq:decay}
\end{equation}

\begin{figure}
\centering
\begin{tabular}{c}
\subfloat[\label{fig:parametric_2D}Two dimensional likelihood contours for the set of 8 cosmological parameters]{\includegraphics[width=0.75\columnwidth]{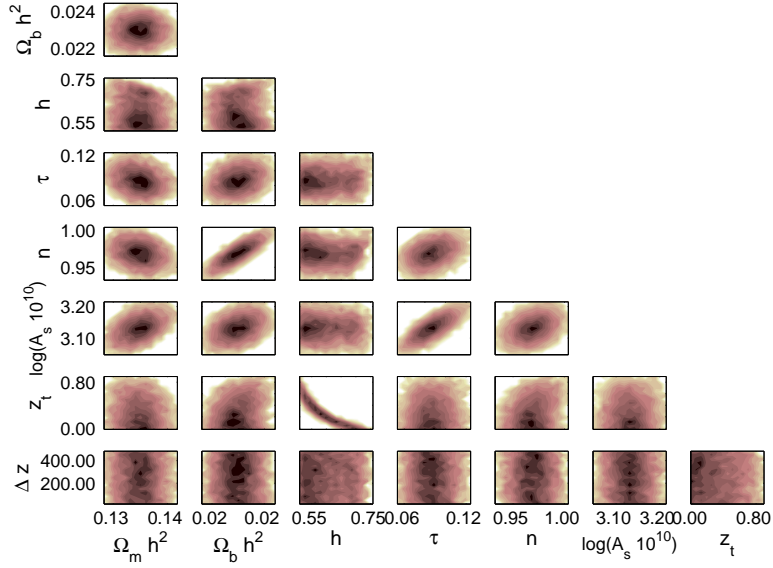}

}
\tabularnewline
\subfloat[\label{fig:parametric_1D}One dimensional marginalized probability distribution]{\includegraphics[width=0.75\columnwidth]{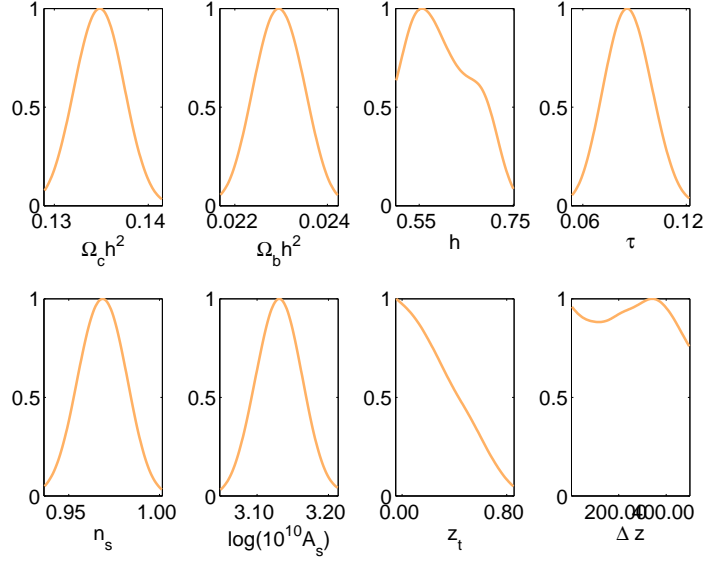}

}
\end{tabular}
\caption{\label{fig:parametric}Cosmological parameter estimation (8 parameters) assuming perturbed decaying dark energy model. $H(z)$ for this model is derived by using the equation of state of dark energy given by Eq.\protect\ref{eq:decay}.}
\end{figure}





\begin{figure}
\centering
\begin{tabular}{lr}
\subfloat[]{\includegraphics[width=0.4\textwidth,height=0.25\textheight]{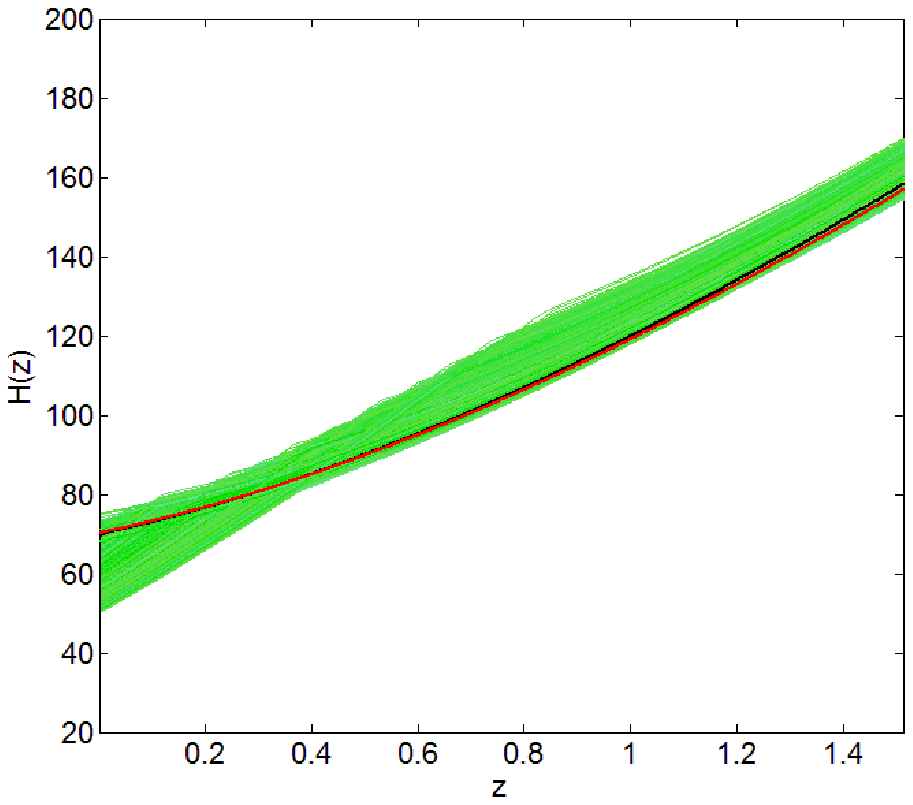}

}
&
\subfloat[]{\includegraphics[width=0.4\textwidth,height=0.25\textheight]{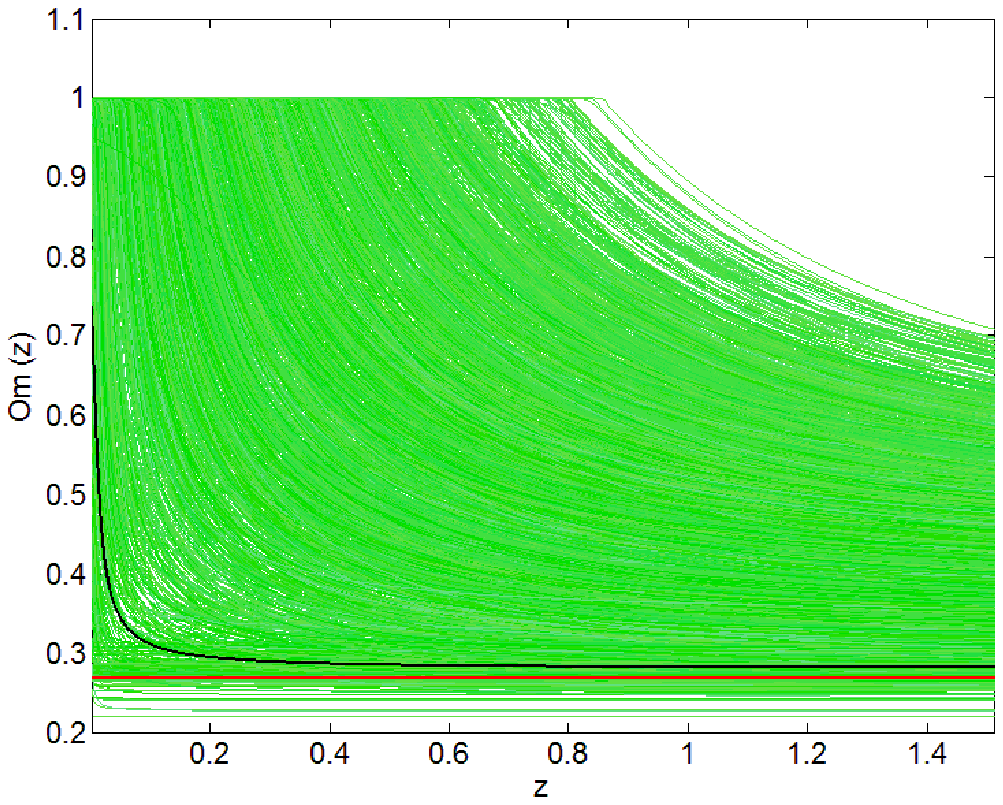}

}
\end{tabular}






\caption{\label{fig:Omhomega8} $95\%$ confidence limit of $H(z)$ and $Om(z)$ for the perturbed decaying dark energy model. $H(z)$ for this model is derived by using the equation of state of dark energy given by Eq.\protect\ref{eq:decay}. Black line is the best fit decaying dark energy model and the red line is the best fit $\Lambda$CDM model.}
\end{figure}

This simple parametric form allows equation of state of dark energy
to change rapidly at low redshifts while at higher redshifts it behaves
exactly similar to $\Lambda$CDM model. It is also bounded to give
$-1<w(z)<0$ not to violate physical laws where it is in fact very
difficult to explain theoretically any phenomena with $w(z)<-1$.
Here $z_{t}$ and the $\Delta z$ are two parameters for the equation
of state of dark energy. The first study set is carried out using
8 parameters amongst which six are the standard cosmological parameters
and two are the equation of state parameters. We have considered a
perturbed dark energy model for this case. Results are shown
in Fig.\ref{fig:parametric_2D} and Fig.\ref{fig:parametric_1D}.
The plot in Fig.\ref{fig:parametric_2D} shows that there is
a negative correlation between the Hubble parameter and the $z_{t}$.
Except that, the correlation between $z_{t}$ and $\Delta z$ with
any other cosmological parameter is very weak. The likelihood surfaces
of $z_{t}$ and $\Delta z$ are very flat, therefore the standard
deviation in these parameters are very high. The estimated cosmological parameters are given in table \ref{tab:Likelihood-1}. In Fig.\ref{fig:Omhomega8}
we have plotted $H(z)$ and $Om(z)$~\cite{Om} 
\footnote{$Om(z)$ is given by $Om(z)=\frac{h(z)^{2}-1}{(1+z)^{3}-1}$.
} with their $95\%$ confidence limits. We can see that decaying dark energy
models can indeed have a consistent fit to the CMB data as well. Though $z_{t} = 0$ is much more probable than other values, one can see that even $z_{t}=0.8$ can not be ruled out with a very high confidence. This reflects the large degeneracy between theoretical models
fitting CMB data.

\begin{figure}
\centering
\begin{tabular}{lr}
\subfloat[\label{fig:parametric_2D-1}Two dimensional likelihood for the set
of 5 cosmological parameters]{\includegraphics[width=0.45\textwidth]{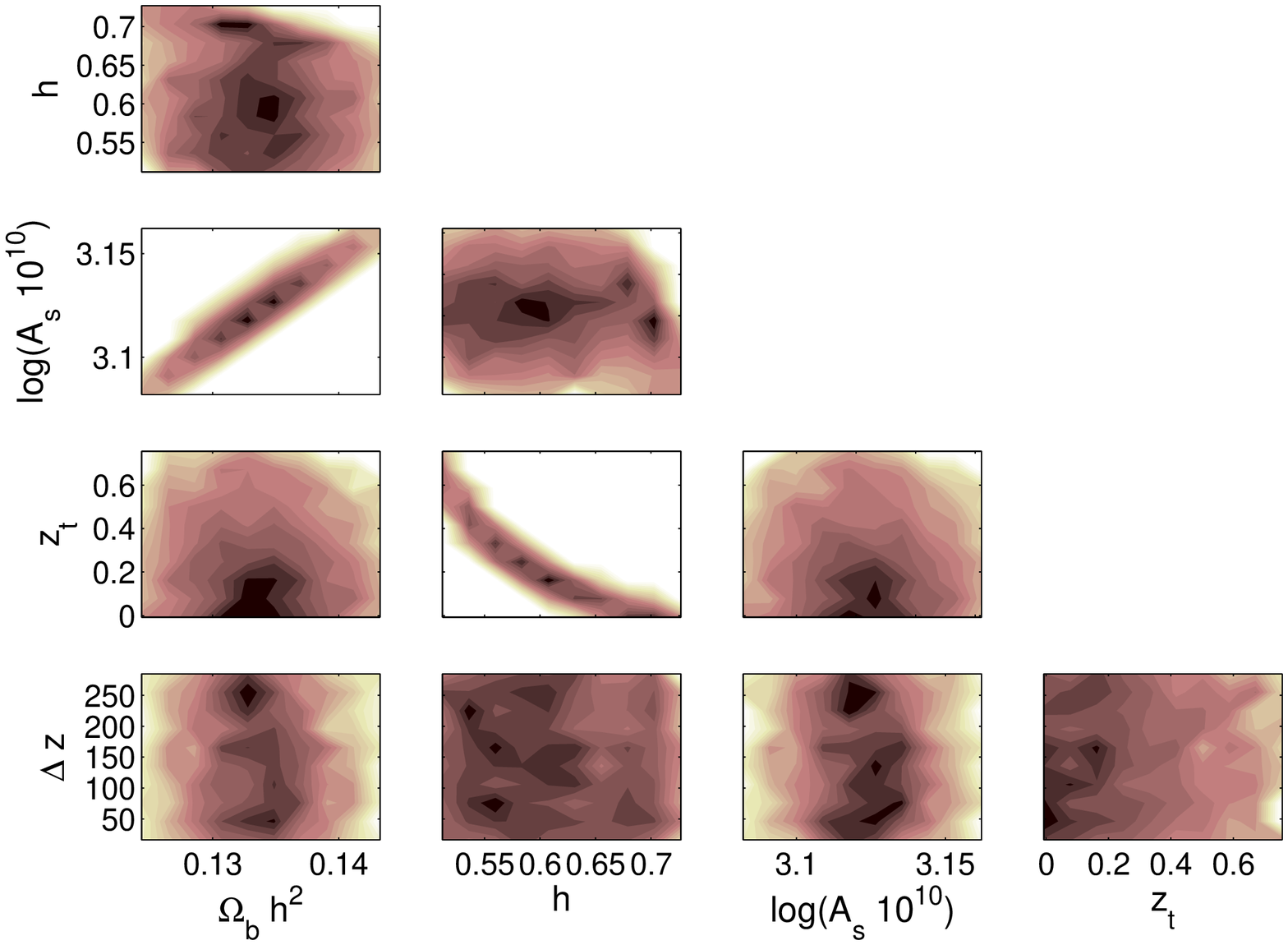}

}
&
\subfloat[\label{fig:parametric_1D-1}One dimensional marginal probability distribution]{\includegraphics[width=0.45\textwidth]{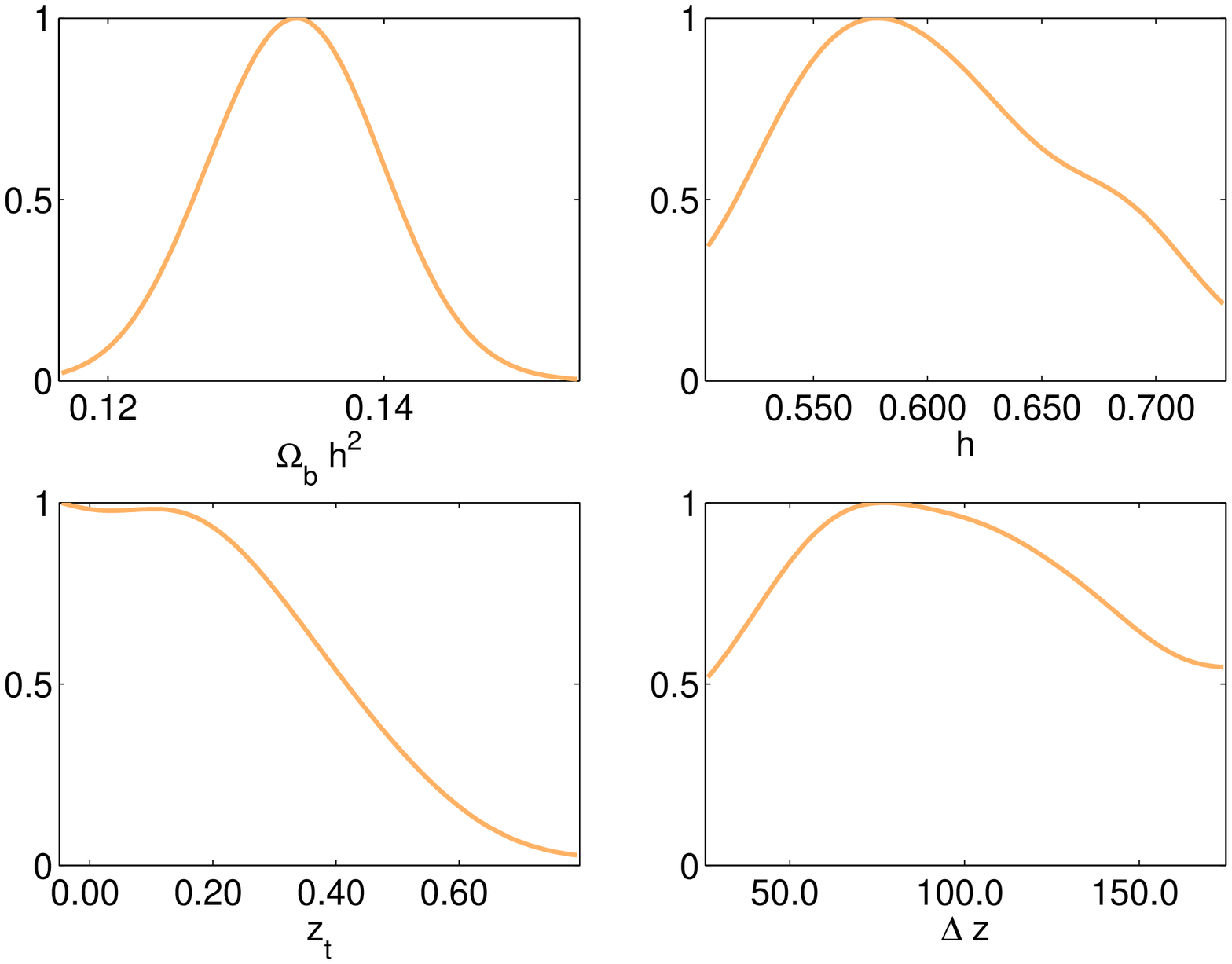}

}\\
\end{tabular}

\caption{\label{fig:parametric-1} Parameter estimation for the perturbed decaying dark energy model. $H(z)$ for this model is derived by using the equation of state of dark energy given by Eq.\protect\ref{eq:decay} and other cosmological parameters are fixed at their best fit $\Lambda$CDM values.}
\end{figure}

The second analysis is carried out using 5 parameters namely $\Omega_{0m}h^{2}$,
$h$, $A_{s}$, $z_{t}$ and $\Delta z$. We have considered a perturbed
dark energy model for this case as well. Results are shown in
Fig.\ref{fig:parametric_2D-1} and Fig.\ref{fig:parametric_1D-1}.
The likelihood contours of $z_{t}$ and $\Delta z$ are very much
flat and constraints on $z_{t}$ and $\Delta z$ are slightly tighter
than the case of 8 parameter analysis. 
The estimated cosmological parameters are given in table \ref{tab:Likelihood-1}.






\begin{table}
\footnotesize
\begin{tabular}{|c|c|c|c|c|}
\hline 
 & \multicolumn{1}{c}{8 parameters} &  & \multicolumn{1}{c}{5 parameters} & \tabularnewline
\hline 
 & Average  & Max-likelihood  & Average  & Max-likelihood\tabularnewline
\hline
\hline 
$\Omega_{b}h^{2}$  & 0.0224$\pm$0.0005  & 0.0224  & 0.0224  & \tabularnewline
\hline 
$\Omega_{0m}h^{2}$  & 0.1343$\pm$0.0055  & 0.1357  & 0.1336$\pm$0.0048  & 0.1331\tabularnewline
\hline 
$h$  & 0.6136$\pm$0.0629  & 0.6955  & 0.6033$\pm$0.0584  & 0.6990\tabularnewline
\hline 
$\tau$  & 0.0874$\pm$0.0147  & 0.0849  & 0.0848  & \tabularnewline
\hline 
$n_{s}$  & 0.9698$\pm$0.0140  & 0.9676  & 0.968  & \tabularnewline
\hline 
$\log(10^{10}A_{s})$  & 3.1329$\pm$0.0349  & 3.1340  & 3.1213$\pm$0.0202  & 3.1199\tabularnewline
\hline 
$z_{t}$  & 0.2874$\pm$0.2300  & 0.0029  & 0.2226$\pm$0.1885  & -0.0182\tabularnewline
\hline 
$\Delta z$  & 256.4648$\pm$140.6126  & 119.3598  & 141.9368$\pm$81.4316  & 48.1929\tabularnewline
\hline 
$\Delta\chi^{2}$  &  & 1.3  &  & 0.7\tabularnewline
\hline 
\end{tabular}\caption{\label{tab:Likelihood-1} Parameter estimation for the decaying dark energy model. $H(z)$ for this model is derived by using the equation of state of dark energy given by Eq.\protect\ref{eq:decay}.}
\end{table}

It is worth mentioning that the best fit parameters for the $\Lambda$CDM
model are $\Omega_{b}h^{2}=0.0224$, $\Omega_{0m}h^{2}=0.1336$,
$h=0.7097$ , $\tau=0.0848$, $n_{s}=0.970$. The best fit parameters 
for the decaying dark energy model show an improvement
of $\Delta\chi^{2}=1.3$ with respect to $\Lambda$CDM model. This
improvement is not significant to consider this model being favored
to the standard model but considering the fact that this model also
has a better fit to the supernovae and BAO data in comparison to the
the standard model, this finding might be interesting. Looking at
Fig.\ref{fig:parametric_1D} and Fig.\ref{fig:Omhomega8} we can
see another interesting result that assuming a decaying dark energy
model allows much lower values of $H_{0}$ to become consistent to
the CMB data.


\section{Conclusion}

The analysis has been carried out by doing some analytical calculations
estimating the effects of the changes in the expansion history of
the Universe on the CMB angular power spectrum. Motivated by our analytical
analysis we considered two different forms of parametrizations for
the expansion history and we studied the effect on the CMB low multipoles
and ISW plateau using MCMC analysis. First parametric form assumes
a Gaussian bump in addition to the $H(z)$ given by the standard $\Lambda$CDM
model. Assuming this parametric form allow us to study how far we
can deviate from $\Lambda$CDM model and still having a concordance
to the data. This parametrisation also can help us to test the standard
model itself and look for any possible deviation favoured by the data.
Our analysis shows that it is possible to deviate significantly from
the $H(z)$ given by the $\Lambda$CDM model at some intermediate
redshift ranges (it behaves more as an unbound parameter) and still having a proper fit to the data. Our analysis
also shows that the spatially flat $\Lambda$CDM model is in proper
concordance to the data and this model stands close to the centre
of confidence contours. In the second parametric form we considered
a decaying dark energy model at low redshifts. Our analytical and
intuitive calculations indicates that increases in the expansion history
at low redshifts might result to a better fit to the data. This is
a feature of decaying dark energy models where we observe a slowing
down of the acceleration in the expansion history of the Universe.
Indeed we realised that this parametric form can result to a better
fit up to $\Delta\chi^{2}=1.3$ in comparison with the best fit $\Lambda$CDM
model. This result indicate this slowing down model is not ruled out
in favor of $\Lambda$CDM model by its CMB ISW effect. Considering
the fact that these decaying dark energy models also are favoured
mildly by supernovae and BAO observations, makes this finding interesting
and worth further investigation. Another important outcome of our
analysis, is that assuming different parametric forms of $H(z)$ results
in significant changes in the posteriors of the cosmological parameters. For instance assuming the decaying dark energy model, $H_{0}$ can
hold relatively smaller values than those expected from $\Lambda$CDM
model and still having a good fit to the CMB data. This is another
important issue in estimation of cosmological parameters using CMB
data since it is also known that assuming different forms of the primordial
spectrum affects significantly on the estimation of cosmological parameters~\cite{shafieloo_souradeep2011,hazra_shafieloo_souradeep_prd2013}.
The recent results from Planck \cite{PlanckXV} also suggest the deficit of power at
low multipole ($\ell<30$) is a feature of the CMB sky that may be
worth addressing via ISW effect.


\acknowledgments{We would like to thank Eric Linder, Dhiraj Hazra and Stephen Appleby for useful discussions. S.D. acknowledge the Council of Scientific and Industrial Research (CSIR), India for financial support through Senior Research fellowships. Computations were carried out at the HPC facilities at IUCAA. A.S. acknowledge the Max Planck Society (MPG), the Korea
Ministry of Education, Science and Technology (MEST), Gyeongsangbuk-Do and Pohang City for the support of the Independent Junior Research Groups at the Asia Pacific Center for Theoretical Physics (APCTP). T.S. acknowledges Swarnajayanti fellowship grant of DST India.}

\end{document}